# Cures for the Expansion Shock and the Shock Instability of the Roe Scheme

Xue-song Li*, Xiao-dong Ren, Chun-wei Gu

*Key Laboratory for Thermal Science and Power Engineering of Ministry of Education, Department of Thermal Engineering, Tsinghua University, Beijing 100084, PR China*

A common defect of the Roe scheme is the production of non-physical expansion shock and shock instability. An improved method with several advantages was presented to suppress the shock instability. However, this method cannot prevent expansion shock and is incompatible with the traditional curing method for expansion shock. Therefore, the traditional curing mechanism is analyzed. The discussion explains the effectiveness of the traditional curing method and identifies several defects, one of which leads to incompatibility between curing the shock instability and expansion shock. Consequently, a new improved Roe scheme is proposed in this study. This scheme is concise, easy to implement, low computational cost, and robust. More importantly, the scheme can simultaneously cure the shock instability and expansion shock without additional costs.

**Key word:** Roe scheme, Expansion shock, Shock instability

## 1. Introduction

The Roe scheme [1] is one of the most famous and important shock-capturing schemes because of its high accuracy. This scheme has undergone considerable development, such as its extension to incompressible flows [2][3][4], and has been extensively used for flow computation, such as in Euler flows [5], LES [6][7], and cavitation [8]. However, the Roe scheme also suffers from a few shortcomings, such as shock instability and expansion shock [9].

* Corresponding author. Tel.: 0086-10-62794617; fax: 0086-10-62795946
E-mail address: xs-li@mail.tsinghua.edu.cn (X.-S. Li).

Shock instability is a well-known defect of supersonic flows with different performances, such as carbuncle, kinked Mach stem, and odd–even decoupling. Several methods were proposed to cure shock instability; the cures were achieved by adding an entropy fix [10], combining a dissipative scheme [9], increasing the basic upwind dissipation [11][12], and considering multi-dimensional characteristics [13][14].

The expansion shock is another defect of the Roe scheme, which is an unphysical solution that violates the entropy condition. Moreover, this defect often yields unacceptable values, such as negative pressure and density, and leads to the divergence of computation for the highly energetic flow. The entropy fix is often adopted to overcome this drawback, but this approach has limited effects while introducing large numerical dissipation and unfavorable empirical parameters. Another considerably common curing method introduces a slight modification by redefining the numerical signal velocities with improved results [12][15][16].

In Ref. [17], the momentum interpolation mechanism in the Roe scheme [18][19][20] is considered the most important reason for shock instability. Thus, a new improved Roe scheme is proposed [17] by removing the momentum interpolation mechanism for the non-linear flow. This improvement cures the shock instability while removing the problem-independence empirical parameters and decreasing the numerical dissipation. However, in this paper several numerical results show an unexpected defect, wherein the expansion shock becomes serious and the traditional curing method is invalid. To broaden the range of applications of the improved Roe scheme, the current study aims to cure the expansion shock by identifying the reason for the deterioration and further elucidating the mechanism that produces the expansion shock, and thus propose an ideal scheme.

The rest of this paper is organized as follows. Chapter 2 provides the governing equations and the improved Roe scheme for curing shock instability [17]. Chapter 3 analyzes the mechanism of the traditional method of curing the expansion shock. Chapter 4 provides a new approach to cure the expansion shock while maintaining all the advantages of the improved Roe scheme. Chapter 5 concludes this paper.

## 2. Governing Equations and the Roe Scheme

### 2.1 Governing Equations

The governing three-dimensional Navier–Stokes equations can be written as follows:

$$\frac{\partial \boldsymbol{Q}}{\partial t}+\frac{\partial \boldsymbol{F}}{\partial x}+\frac{\partial \boldsymbol{G}}{\partial y}+\frac{\partial \boldsymbol{H}}{\partial z}=0, \tag{1}$$

where $\boldsymbol{Q}=\begin{bmatrix}\rho\\ \rho u\\ \rho v\\ \rho w\\ \rho E\end{bmatrix}$ is the vector of the conservation variables; $\boldsymbol{F}=\begin{bmatrix}\rho u\\ \rho u^2+p\\ \rho uv\\ \rho uw\\ \rho uH\end{bmatrix}$,

$\boldsymbol{G}=\begin{bmatrix}\rho v\\ \rho uv\\ \rho v^2+p\\ \rho vw\\ \rho vH\end{bmatrix}$, and $\boldsymbol{H}=\begin{bmatrix}\rho w\\ \rho uw\\ \rho vw\\ \rho w^2+p\\ \rho wH\end{bmatrix}$ are the vectors of the Euler fluxes; $\rho$ is the fluid density; $p$ is the pressure; $E$ is the total energy; $H$ is the total enthalpy; $u$, $v$, and $w$ are the velocity components in the Cartesian coordinates $(x, y, z)$, respectively.

## 2.2 Roe Scheme and Improvement

The classical Roe scheme can be expressed in the following general form as the sum of a central term and a numerical dissipation term:

$$\tilde{\boldsymbol{F}} = \tilde{\boldsymbol{F}}_{\boldsymbol{c}} + \tilde{\boldsymbol{F}}_{\boldsymbol{d}}, \qquad (2)$$

where $\tilde{\boldsymbol{F}}_{\boldsymbol{c}}$ is the central term and $\tilde{\boldsymbol{F}}_{\boldsymbol{d}}$ is the numerical dissipation term. For a cell face of the finite volume method,

$$\tilde{\boldsymbol{F}}_{\boldsymbol{c},1/2} = \frac{1}{2}\left(\bar{\boldsymbol{F}}_L + \bar{\boldsymbol{F}}_R\right), \qquad (3)$$

$$\bar{\boldsymbol{F}} = U\begin{bmatrix}\rho \\ \rho u \\ \rho v \\ \rho w \\ \rho H\end{bmatrix} + p\begin{bmatrix}0 \\ n_x \\ n_y \\ n_z \\ 0\end{bmatrix}, \qquad (4)$$

where $n_x$, $n_y$, and $n_z$ are the components of the face-normal vector, and $U = n_x u + n_y v + n_z w$ is the normal velocity on the cell face.

According to Ref. [21], a scale uniform framework for the shock-capturing scheme is proposed [22]. This framework is simple and easy to analyze and improve with low computation cost.

$$\tilde{\boldsymbol{F}}_{\boldsymbol{d}} = -\frac{1}{2}\left\{\xi\begin{bmatrix}\Delta\rho \\ \Delta(\rho u) \\ \Delta(\rho v) \\ \Delta(\rho w) \\ \Delta(\rho E)\end{bmatrix} + \left(\delta p_u + \delta p_p\right)\begin{bmatrix}0 \\ n_x \\ n_y \\ n_z \\ U\end{bmatrix} + \left(\delta U_u + \delta U_p\right)\begin{bmatrix}\rho \\ \rho u \\ \rho v \\ \rho w \\ \rho H\end{bmatrix}\right\}, \qquad (5)$$

where the first term on the right side $\xi$ is the basic upwind dissipation; the term $\delta p_p$ is the pressure-difference-driven modification for the cell face pressure, $\delta p_u$ is the velocity-difference-driven modifications for the cell face pressure, $\delta U_u$ is the velocity-difference-driven modification for the cell face velocity, and $\delta U_p$ is the pressure-difference-driven modification for the cell face velocity.

For the classical Roe scheme,

$$\xi = \lambda_1 , \tag{6}$$

$$\delta p_u = \left( \frac{\lambda_5 + \lambda_4}{2} - \lambda_1 \right) \rho \Delta U , \tag{7}$$

$$\delta p_p = \frac{\lambda_5 - \lambda_4}{2} \frac{\Delta p}{c} , \tag{8}$$

$$\delta U_u = \frac{\lambda_5 - \lambda_4}{2} \frac{\Delta U}{c} , \tag{9}$$

$$\delta U_p = \left( \frac{\lambda_5 + \lambda_4}{2} - \lambda_1 \right) \frac{\Delta p}{\rho c^2} , \tag{10}$$

where $c$ is the sound speed, and the eigenvalues of the system are defined as follows:

$$\lambda_1 = \lambda_2 = \lambda_3 = |U| , \tag{11}$$

$$\lambda_4 = |U - c| , \tag{12}$$

$$\lambda_5 = |U + c| . \tag{13}$$

Based on Eqs. (11)–(13), Eqs. (7)–(10) can also be further simplified as:

$$\delta p_u = \max\left(0, c - |U|\right) \rho \Delta U , \tag{14}$$

$$\delta p_p = \operatorname{sign}(U) \min\left(|U|, c\right) \frac{\Delta p}{c} , \tag{15}$$

$$\delta U_u = \operatorname{sign}(U) \min\left(|U|, c\right) \frac{\Delta U}{c} , \tag{16}$$

$$\delta U_p = \max\left(0, c - |U|\right) \frac{\Delta p}{\rho c^2} . \tag{17}$$

Ref. [17] suggests that the shock instability is mainly attributed to the term $\delta U_p$

and can be cured by multiplying Eq. (17) to the two functions $s_1$ and $s_2$ as follows:

$$\delta U_p = s_1 s_2 \max\left(0, c - |U|\right)\frac{\Delta p}{\rho c^2}, \tag{18}$$

$$s_1 = 1 - \left[\min\left(M\frac{\sqrt{4+\left(1-M^2\right)^2}}{1+M^2}, 1\right)\right]^8, \tag{19}$$

where $M$ is the Mach number. The function $s_2$, which is a shock detector and can be obtained in Ref. [17], is not presented in this paper because it is relatively complicated and probably unnecessary for general cases. This improvement is simple and effective.

### 2.3 Two Classical Numerical Tests

Two classical numerical examples are available to test the expansion shock by the proposed scheme. One is a specific one-dimensional shock tube and the other is the supersonic corner problem. The initial condition of the shock tube is given as $\rho_L = 3$, $u_L = 0.9$, $p_L = 3$, $\rho_R = 1$, $u_R = 0.9$, and $p_R = 1$ at the *x*-axis position of 0.3. The supersonic corner problem considers a moving supersonic shock around a 90° corner. The initial condition is given as $\rho_L = 7.04108$, $u_L = 4.07794$, $v_L = 0$, $p_L = 30.05942$, $\rho_R = 1.4$, $u_R = v_R = 0$, and $p_R = 1$ at the *x*-axis position of 0.05. In this study, the mesh grids are 200 for the shock tube, and 400*400 for the supersonic corner. For time discretization, the four-stage Runge–Kutta scheme is adopted. For space discretization, the first-order accuracy is adopted (unless otherwise specified) to discuss the schemes themselves.

## 3. Analysis of the Traditional Method Curing the Expansion Shock

### 3.1 Traditional Curing Method

To avoid the expansion shock, the tradition curing method redefines the physical signal velocities (i.e., non-linear eigenvalues $\lambda_4$ and $\lambda_5$). For example, Ref. [15] proposed that:

$$\lambda_4 = \left|\min\left(U - c, U_L - c_L\right)\right|, \tag{20}$$

$$\lambda_5 = \left|\max\left(U + c, U_R + c_R\right)\right|. \tag{21}$$

To precisely obtain a contact discontinuity, Ref. [12] suggested that only $U$ in $\lambda_4$ and $\lambda_5$ should be improved:

$$\lambda_4 = \left|\min\left(U - c, U_L - c\right)\right|, \tag{22}$$

$$\lambda_5 = \left|\max\left(U + c, U_R + c\right)\right|. \tag{23}$$

Eqs. (22)–(23) can retain the capability of Eqs. (20)–(21) to suppress the expansion shock while having the advantage of obtaining the contact discontinuity. Therefore, only Eqs. (22)–(23) are discussed in this study.

### 3.2 Performances of the Schemes

Figs. 1 and 2 show the results by the classical Roe scheme, as described by Eqs. (6) and (14)–(17); and the traditional curing method, as described by Eqs. (6)–(11) and (22)–(23).

For the shock tube, the classical Roe scheme evidently produces an expansion shock at the $x$ = 0.3 position. The traditional curing method demonstrates substantially improved performance. However, a sight gap also exists with the traditional method.

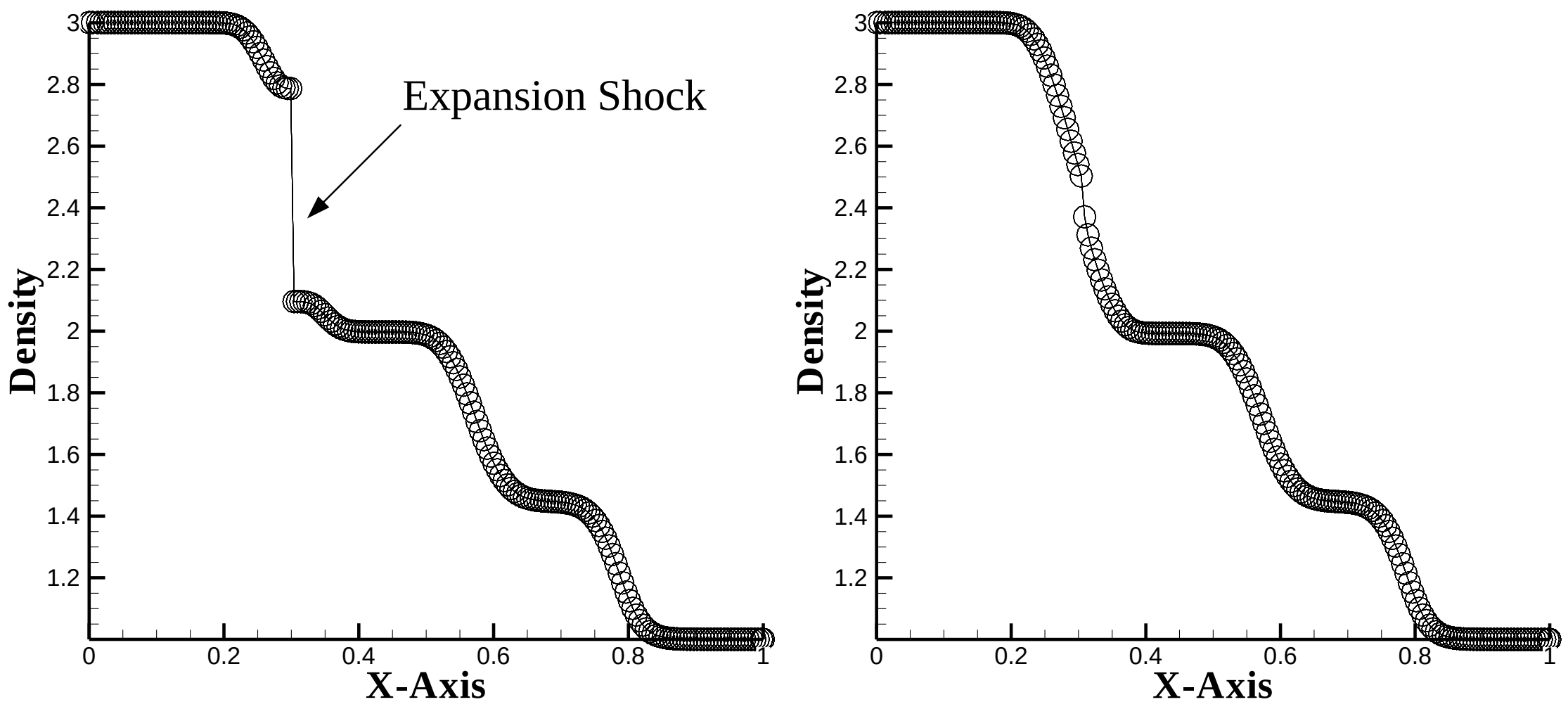

(a) Classical Roe scheme (b) Traditional curing method

Fig. 1 Results of the shock tube test at $t$ = 0.2 s

For the supersonic corner, a series expansion waves exist around the corner. Thus, the numerical computation could produce an expansion shock. The results of the classical Roe scheme are shown in Fig. 2(a). No evident expansion shock was observed, but the shock instability was expectedly strong. The traditional curing method produces similar results (see Fig. 2(b)).

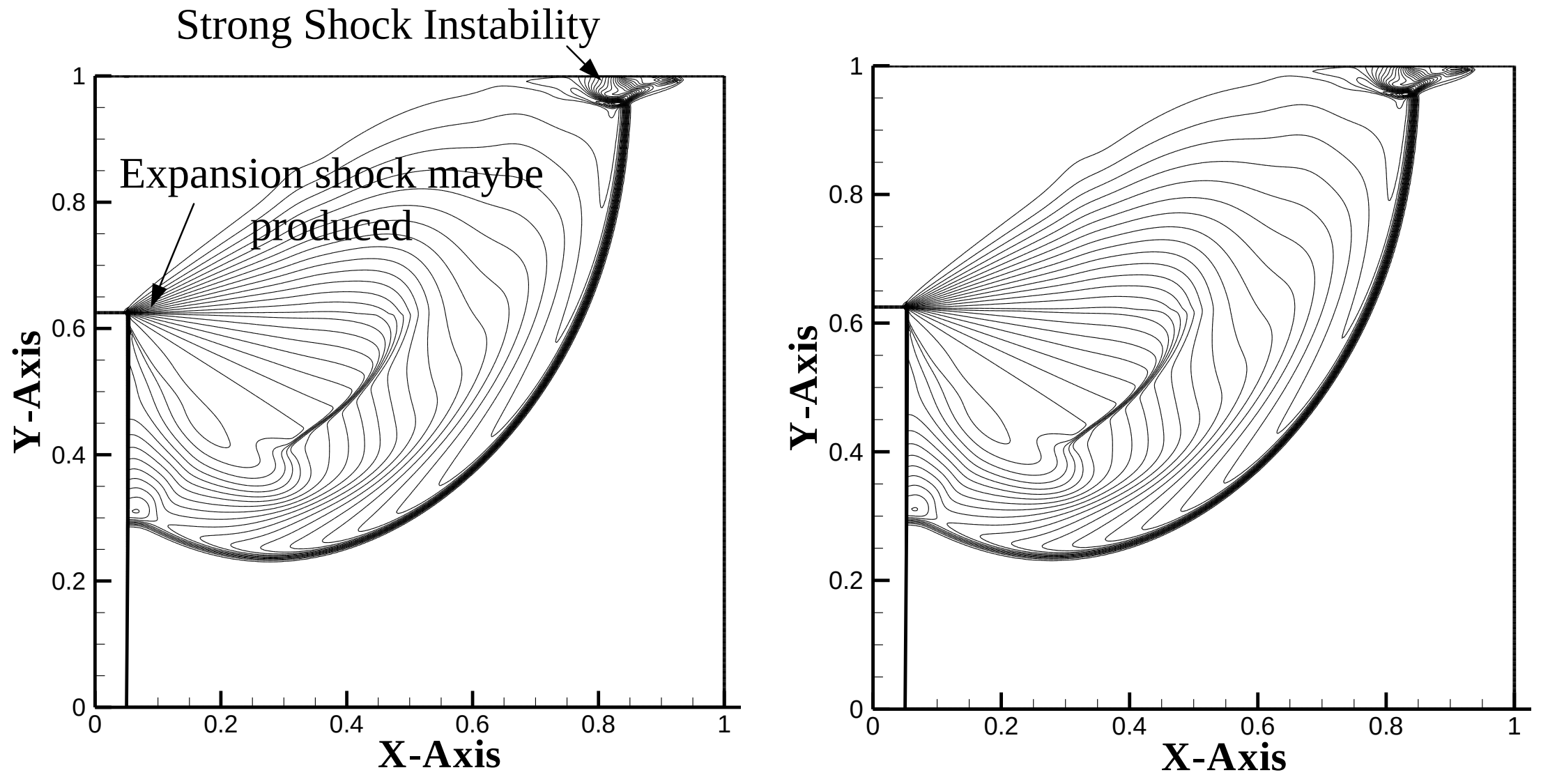

(a) Classical Roe scheme (b) Traditional curing method

Fig. 2 Results of the supersonic corner test at $t$ = 0.155 s

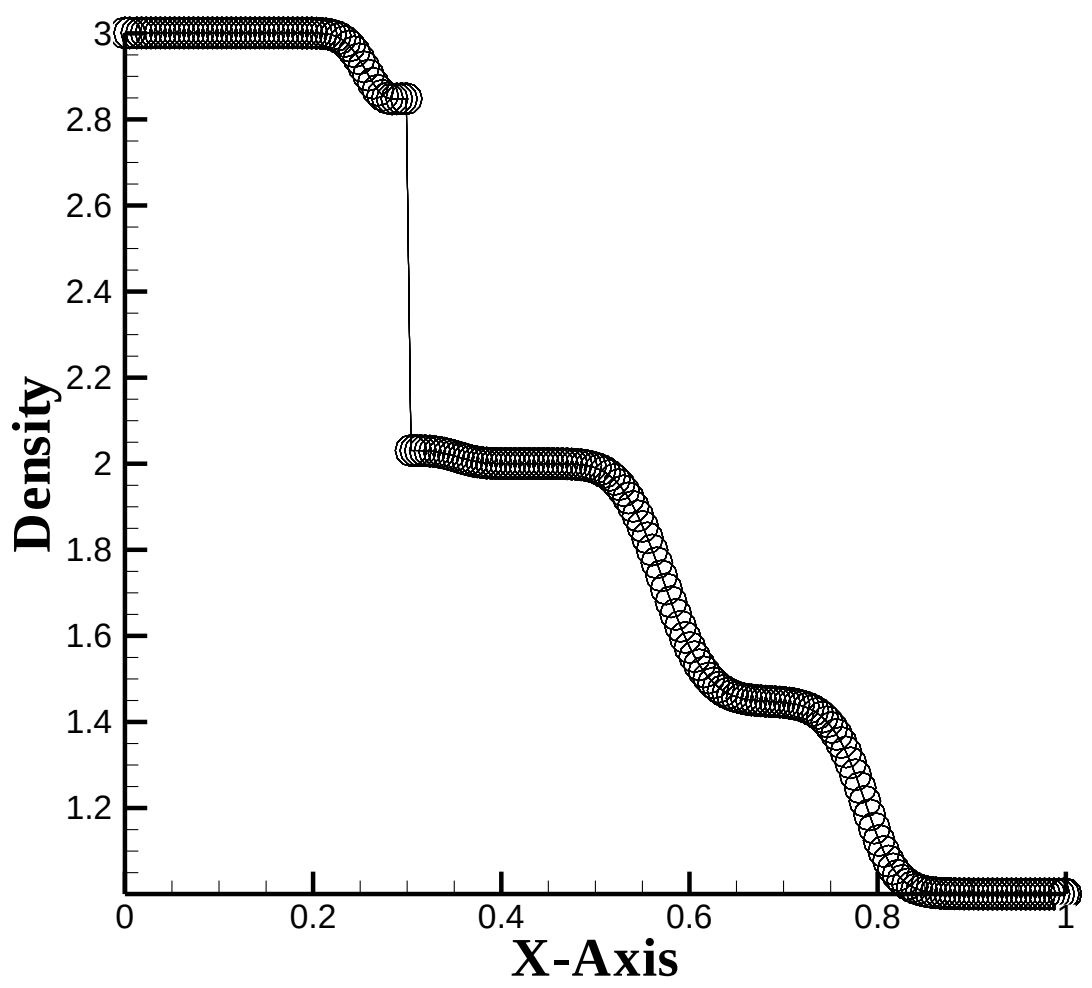

(a) Shock tube

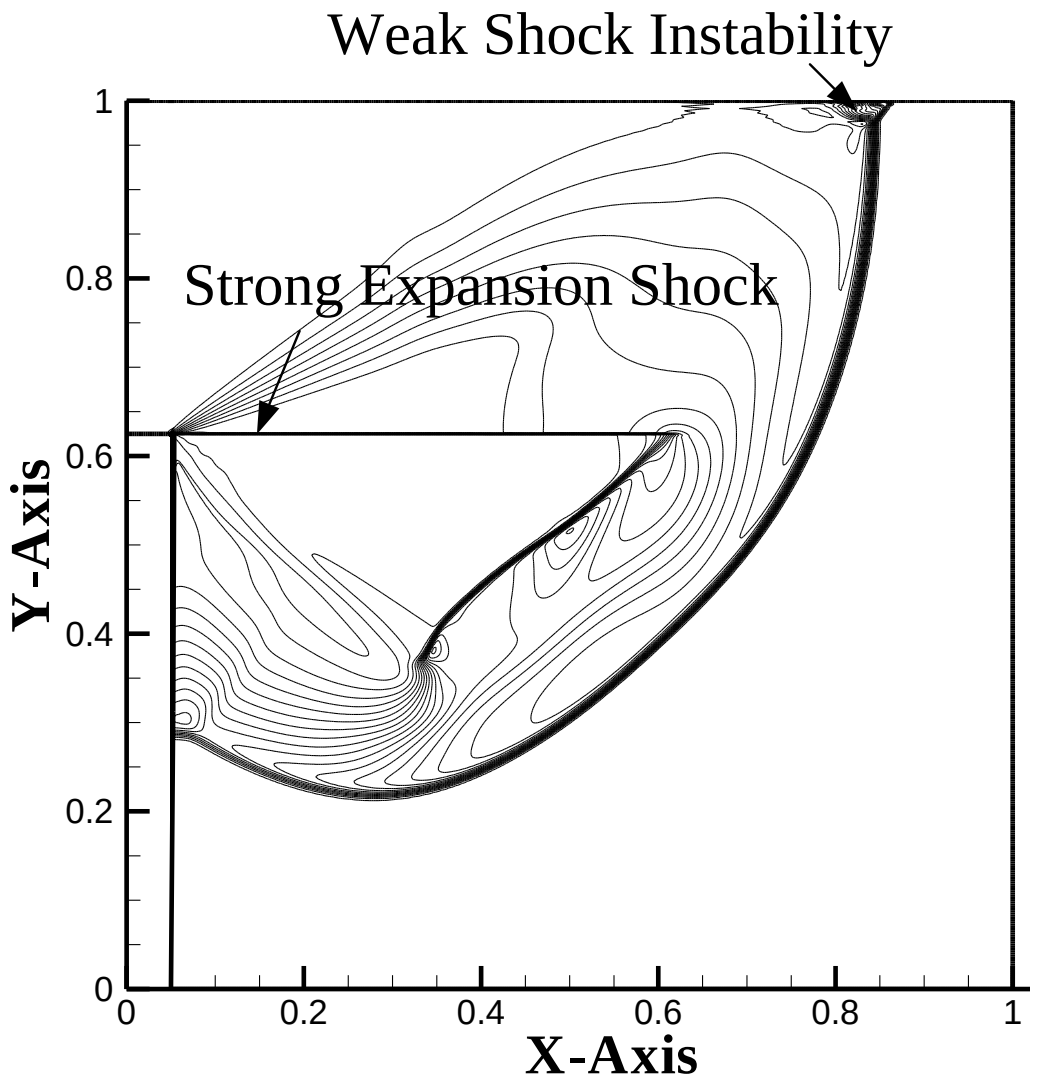

(b) Supersonic corner

Fig. 3 Results by the improved Roe scheme as described by Eqs. (6), (14)–(16), and (18)

By employing the improved Roe scheme in Eqs. (6), (14)–(16) and (18) [17], the results of the one-dimensional shock tube are similar to those with the classical Roe scheme (see Fig. 3(a)) because the improvement of Eq. (18) only affects multi-dimensional computation as analyzed in Ref. [17]. For the two-dimensional computation of the supersonic corner, results become significantly different from those of the classical Roe scheme (see Fig. 3(b)). The shock instability becomes substantially weak and is nearly cured. However, a strong expansion shock occurs. In the iterative calculation process, the density occasionally becomes negative and the following limitation is necessary to prevent computational divergence:

$$\rho_{\text{iter}} = \max\left(\rho_{\text{cal}}, \varepsilon\right), \tag{24}$$

where $\varepsilon$ is a small positive value.

The traditional curing method for expansion shock in Eqs. (22)–(23) can also be integrated into the improved Roe scheme. For the shock tube, this method can produce results as Fig. 1(b). However, this approach is invalid for the supersonic corner and even substantially increases the expansion shock. The computation diverges because of negative density even when using Eq. (24). This unexpected problem seems confusing and may hinder the possible extensive application of the improved Roe scheme because of concerns regarding computational robust. Therefore, in the following sections the mechanism of preventing expansion shock is further analyzed and the new method satisfies the stringent requirement of simultaneously curing the expansion shock and the shock instability without additional costs.

### 3.3 Analysis of the Schemes

To develop the new method, the mechanism of the traditional curing method is first further analyzed. Eqs. (22)–(23) can be decomposed into the following five conditions:

(1) $|U| < c$:

$$\left|\min\left(U-c, U_L-c\right)\right| = \begin{cases} c-U & U_L > U \\ c-U_L & U_L \le U \end{cases} = |U-c| - \min\left(0, U_L-U\right), \tag{25}$$

$$\left|\max\left(U+c, U_R+c\right)\right| = \begin{cases} U+c & U > U_R \\ U_R+c & U \le U_R \end{cases} = |U+c| + \max\left(0, U_R-U\right). \tag{26}$$

(2) $U > c$ and $U_L > c$ and ($U_R > c$ or $U_R < c$):

$$\left|\min\left(U-c, U_L-c\right)\right| = \begin{cases} U-c & U_L > U \\ U_L-c & U_L \le U \end{cases} = |U-c| + \min\left(0, U_L-U\right), \tag{27}$$

$$\left|\max\left(U+c, U_R+c\right)\right| = \begin{cases} U+c & U > U_R \\ U_R+c & U \le U_R \end{cases} = |U+c| + \max\left(0, U_R-U\right). \tag{28}$$

(3) $U > c$ and $U_L < c$ and $U_R > c$:

$$\left|\min\left(U-c,U_L-c\right)\right|=c-U_L=\left|U-c\right|-\left(U+U_L-2c\right), \tag{29}$$

$$\left|\max\left(U+c,U_R+c\right)\right|=U_R+c=\left|U+c\right|+\max\left(0,U_R-U\right). \tag{30}$$

(4) $U<-c$ and $U_R<-c$:

$$\left|\min\left(U-c,U_L-c\right)\right|=\begin{cases}c-U_L & U>U_L\\ c-U & U<U_L\end{cases}=\left|U-c\right|-\min\left(0,U_L-U\right), \tag{31}$$

$$\left|\max\left(U+c,U_R+c\right)\right|=\begin{cases}-U-c & U>U_R\\ -U_R-c & U\le U_R\end{cases}=\left|U+c\right|-\max\left(0,U_R-U\right). \tag{32}$$

(5) $U<-c$ and $U_R>-c$ and $U_L<-c$:

$$\left|\min\left(U-c,U_L-c\right)\right|=c-U_L=\left|U-c\right|-\min\left(0,U_L-U\right), \tag{33}$$

$$\left|\max\left(U+c,U_R+c\right)\right|=\left|U_R+c\right|=\left|U+c\right|+\left(U+U_R+2c\right). \tag{34}$$

By considering

$$\min\left(0,U_L-U\right)=-\max\left(0,U-U_L\right), \tag{35}$$

Eqs. (25)–(34) can be summarized as follows:

$$\left|\min\left(U-c,U_L-c\right)\right|=\left|U-c\right|-2b_L, \tag{36}$$

$$\left|\max\left(U+c,U_R+c\right)\right|=\left|U+c\right|+2b_R, \tag{37}$$

where

$$b_L=\frac{1}{2}\begin{cases}U+U_L-2c & U>c \text{ and } U_L<c \text{ and } U_R>U_L\\ \operatorname{sign}\left(U-c\right)\max\left(0,U-U_L\right) & \text{otherwise}\end{cases}, \tag{38}$$

$$b_R=\frac{1}{2}\begin{cases}U+U_R+2c & U<-c \text{ and } U_R>-c \text{ and } U_R>U_L\\ \operatorname{sign}\left(U+c\right)\max\left(0,U_R-U\right) & \text{otherwise}\end{cases}. \tag{39}$$

Therefore, Eqs. (7)–(10) become:

$$\delta p_u=\left[\max\left(0,c-\left|U\right|\right)+b_R-b_L\right]\rho\Delta U, \tag{40}$$

$$\delta p_p=\left[\operatorname{sign}\left(U\right)\min\left(\left|U\right|,c\right)+b_R+b_L\right]\frac{\Delta p}{c}, \tag{41}$$

$$\delta U_u=\left[\operatorname{sign}\left(U\right)\min\left(\left|U\right|,c\right)+b_R+b_L\right]\frac{\Delta U}{c}, \tag{42}$$

$$\delta U_p = \left[ \max\left(0, c - |U|\right) + b_R - b_L \right] \frac{\Delta p}{\rho c^2} . \quad (43)$$

In the preceding equations, $U$ is the average of $U_L$ and $U_R$, and the reasonable value of $U$ is between the range of $U_L$ and $U_R$:

$$U \in \left[ U_L, U_R \right] . \quad (44)$$

For the general average method, such as a simple average or Roe average:

$$U \approx \frac{U_L + U_R}{2} . \quad (45)$$

For compression flows, $b_R = b_L = 0$ because $U_R < U < U_L$. Therefore, only the expansion flows are considered for analyzing the increment terms $b_R - b_L$ and $b_R + b_L$ in Eqs. (40)–(43):

(1) $|U| < c$

$$b_R - b_L = \frac{1}{2}\left( U_R - U_L \right) > 0 , \quad (46)$$

$$b_R + b_L = \frac{1}{2}\left( U_R - 2U + U_L \right) \approx 0 . \quad (47)$$

Therefore, for the subsonic expansion flows, $\delta p_u$ and $\delta U_p$ are increased by the modification of Eqs. (22)–(23).

(2) $U > c$ and $U_L > c$

$$b_R - b_L = \frac{1}{2}\left( U_R - 2U + U_L \right) \approx 0 , \quad (48)$$

$$b_R + b_L = \frac{1}{2}\left( U_R - U_L \right) > 0 . \quad (49)$$

Consequently, $\delta p_p$ and $\delta U_u$ increase, but these results are unsuitable. For supersonic flows, all increment terms should be zero because of the upwind characteristics.

(3) $U > c$ and $U_L < c$ and $U_R > c$

$$b_R - b_L = \frac{1}{2}\left[\left(U_R - U\right) - \left(U + U_L - 2c\right)\right] \approx c - U_L > 0\,, \tag{50}$$

$$b_R + b_L = \frac{1}{2}\left[\left(U_R - U\right) + \left(U + U_L - 2c\right)\right] \approx U - c > 0\,. \tag{51}$$

Therefore, the increment terms are unsatisfactory because they are not equal to zero and not smooth between Conditions (1) and (2).

The other two conditions of $U < -c$ are not discussed for simplicity because they produce the same conclusions as the conditions of $U > c$.

### 3.4 Further Analysis of the Curing Expansion Shock Mechanism

The preceding discussion reveals a few unsatisfactory features of the traditional curing method in Eqs. (22)–(23). Moreover, the discussion provides clues regarding the mechanism of expansion shock suppression. Two inspirations are obtained as follows.

(1) An increment factor is designed as follows:

$$\Delta s = \frac{\operatorname{sign}\left(U + c\right)\max\left(0, U_R - U_L\right) - \operatorname{sign}\left(U - c\right)\max\left(0, U_R - U_L\right)}{4}\,, \tag{52}$$

where $U - U_L$ in Eq. (38) and $U_R - U$ in Eq. (39) are replaced by $\frac{U_R - U_L}{2}$ to make $\Delta s$ strictly equal to zero for the compression and supersonic flows $|U| > c$. For the subsonic expansion shock, a positive increment same as Eq. (46) is produced.

(2) Eight cases are designed to test the effect of changing the values of $\delta p_u$, $\delta p_p$, $\delta U_u$, and $\delta U_p$:

$$\delta p_u = \left[\max\left(0, c - |U|\right) + \Delta s_{pu}\right]\rho\Delta U\,, \tag{53}$$

$$\delta p_p = \left[\operatorname{sign}\left(U\right)\min\left(|U|, c\right) + \Delta s_{pp}\right]\frac{\Delta p}{c}\,, \tag{54}$$

$$\delta U_u = \left[\operatorname{sign}\left(U\right)\min\left(|U|, c\right) + \Delta s_{uu}\right]\frac{\Delta U}{c}\,, \tag{55}$$

$$\delta U_p = \left[ \max\left(0, c - |U|\right) + \Delta s_{up} \right] \frac{\Delta p}{\rho c^2} . \tag{56}$$

Table 1 Eight test cases

| | $\Delta s_{pu}$ | $\Delta s_{up}$ | $\Delta s_{pp}$ | $\Delta s_{uu}$ |
|---|---|---|---|---|
| Case 1 | $+\Delta s$ | 0 | 0 | 0 |
| Case 2 (divergence) | $-\Delta s$ | 0 | 0 | 0 |
| Case 3 | 0 | $+\Delta s$ | 0 | 0 |
| Case 4 (divergence) | 0 | $-\Delta s$ | 0 | 0 |
| Case 5 (divergence) | 0 | 0 | $+\Delta s$ | 0 |
| Case 6 | 0 | 0 | $-\Delta s$ | 0 |
| Case 7 (divergence) | 0 | 0 | 0 | $+\Delta s$ |
| Case 8 | 0 | 0 | 0 | $-\Delta s$ |

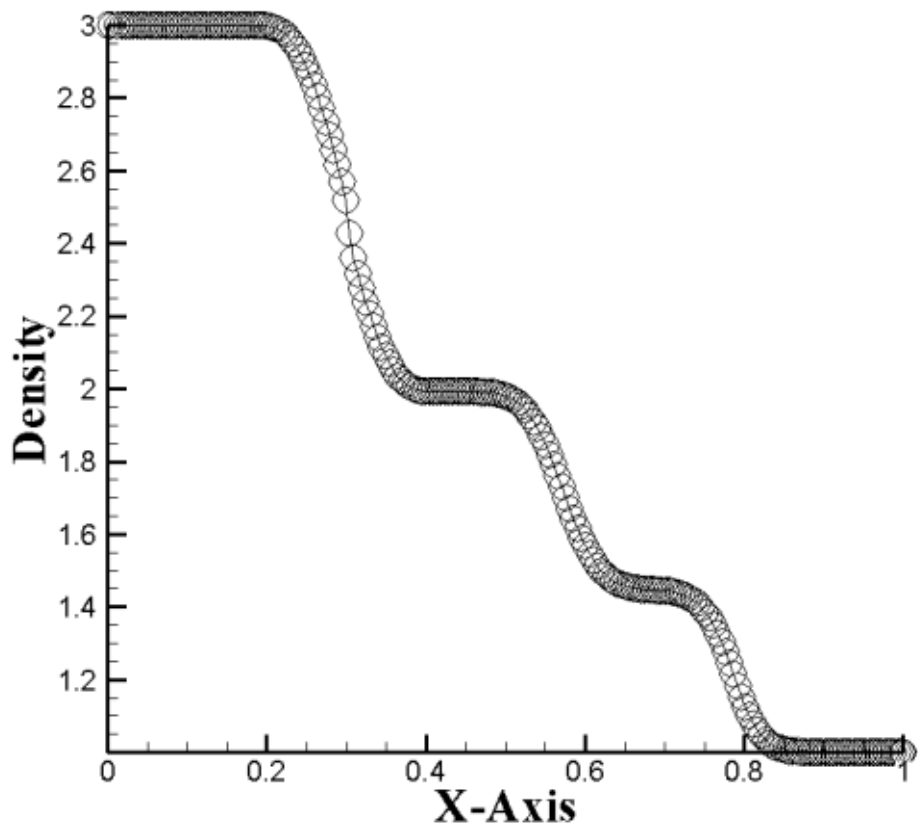

(a) Case 1

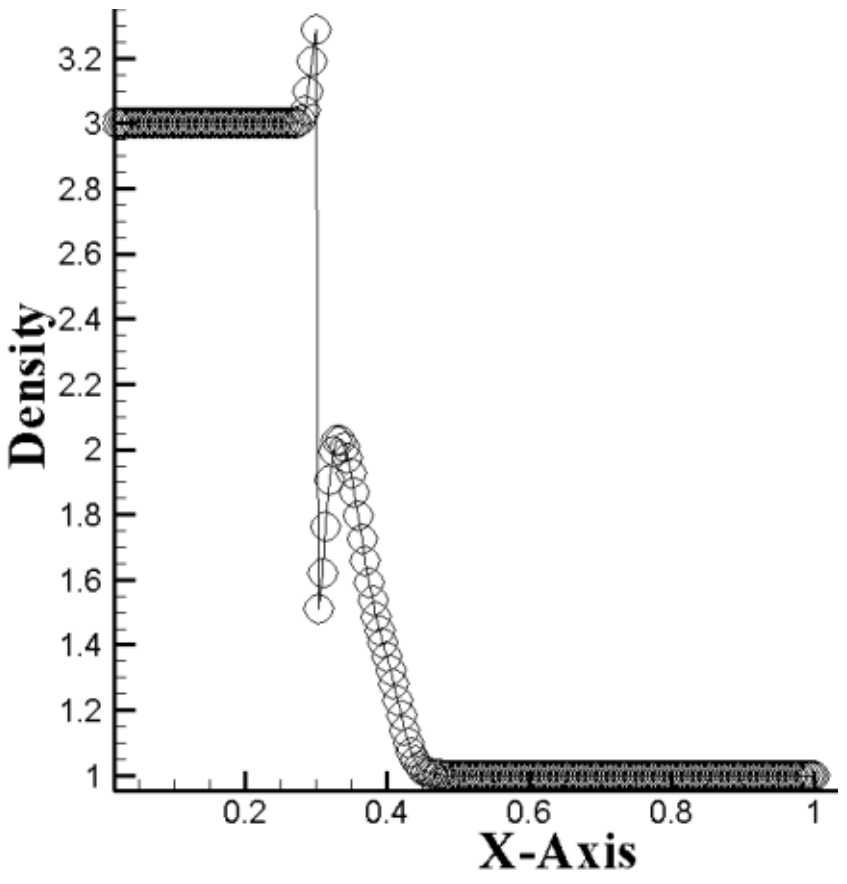

(b) Case 2

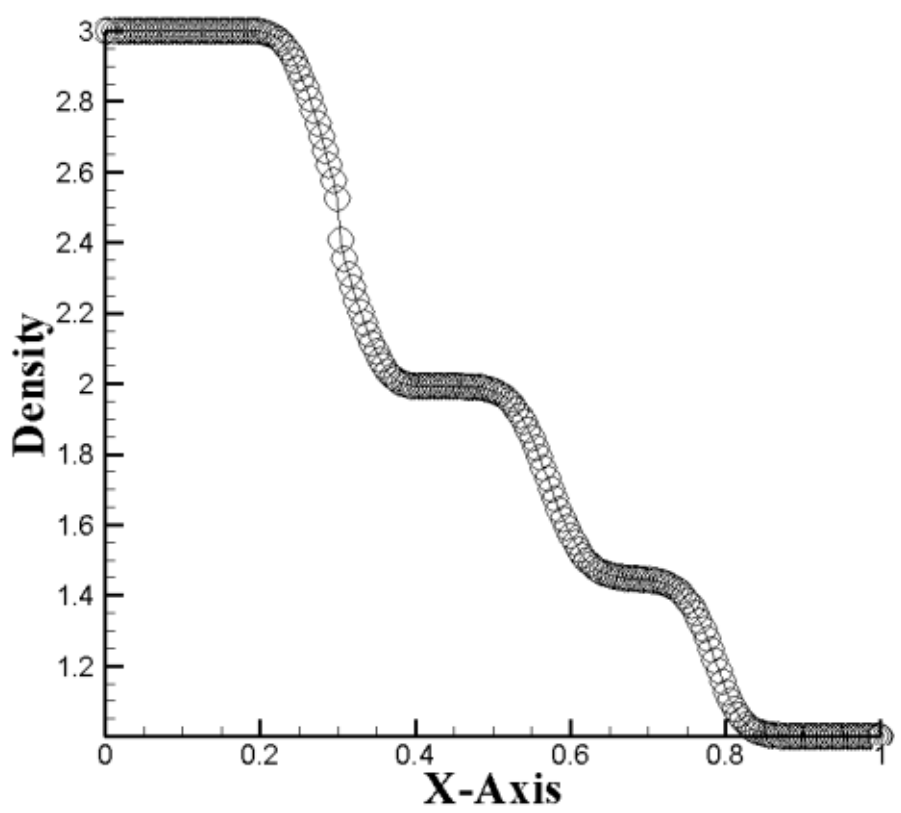

(c) Case 3

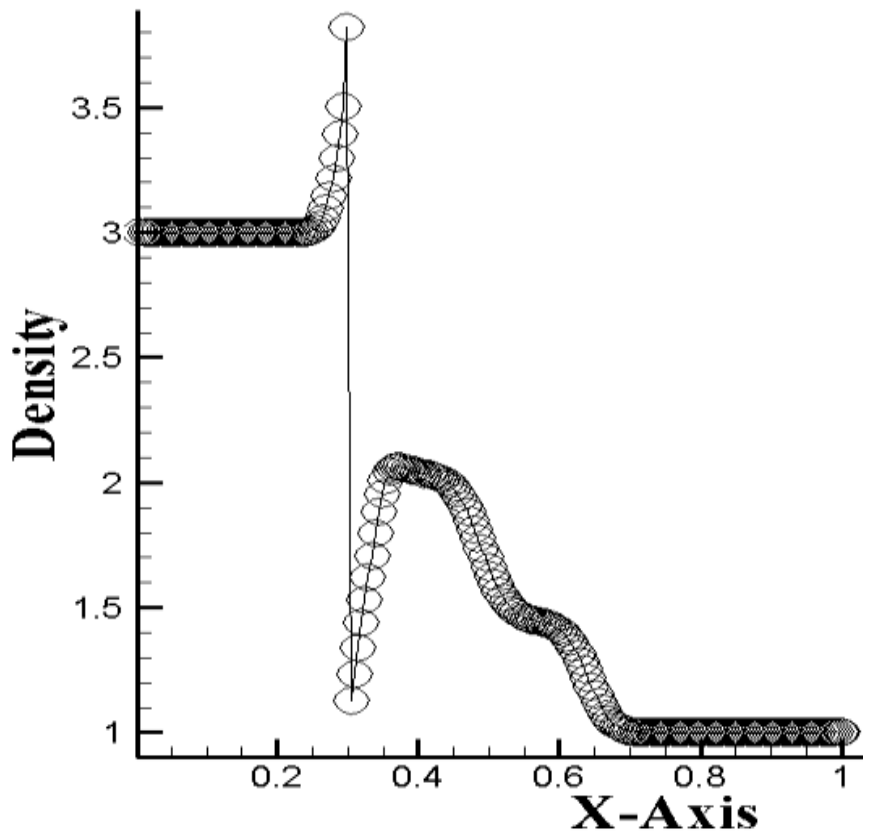

(d) Case 4

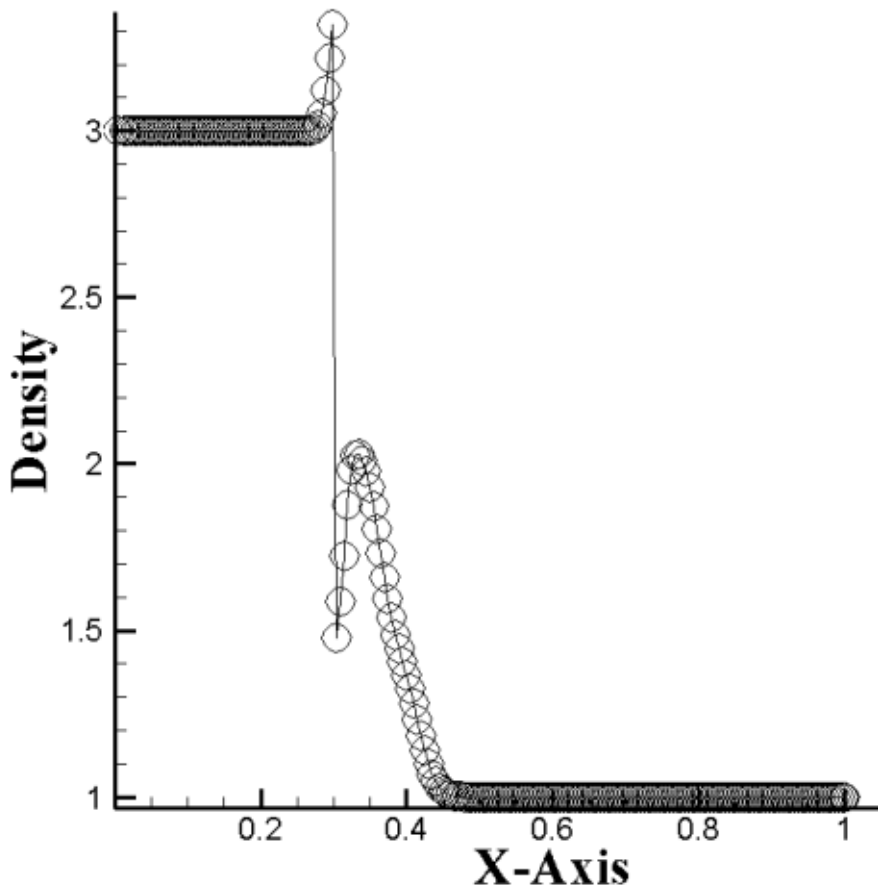

(e) Case 5

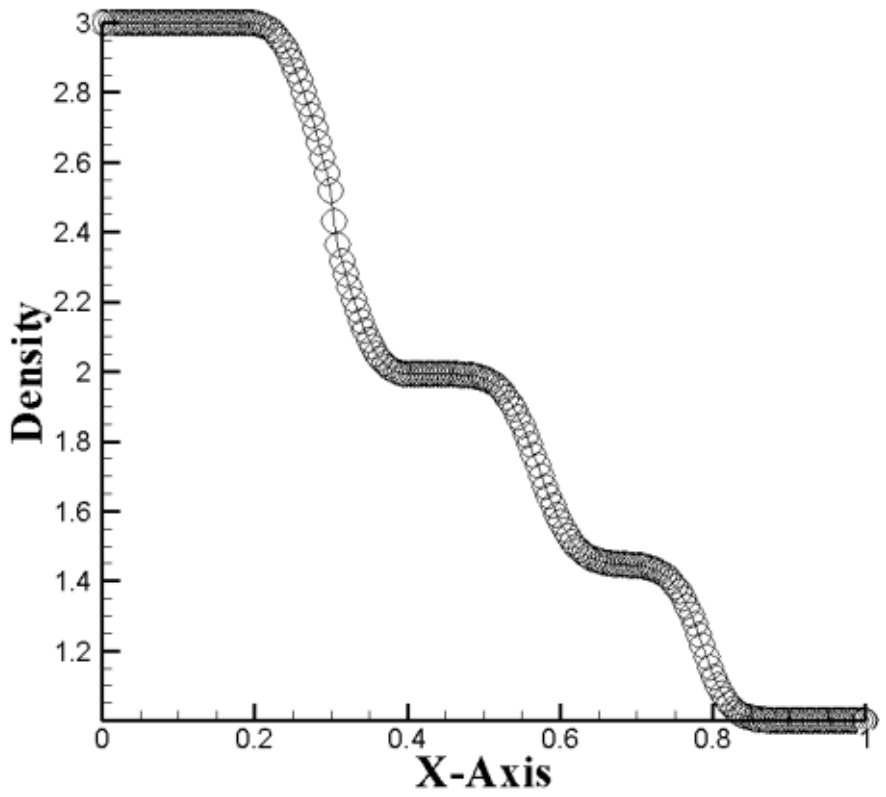

(f) Case 6

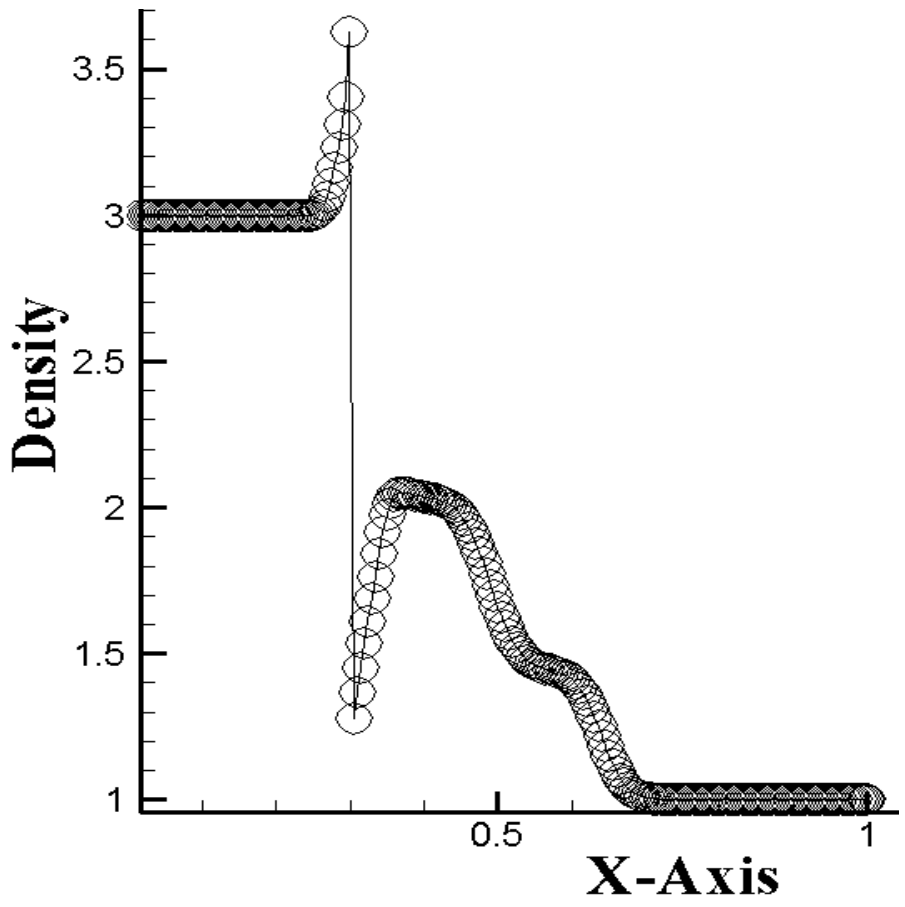

(g) Case 7

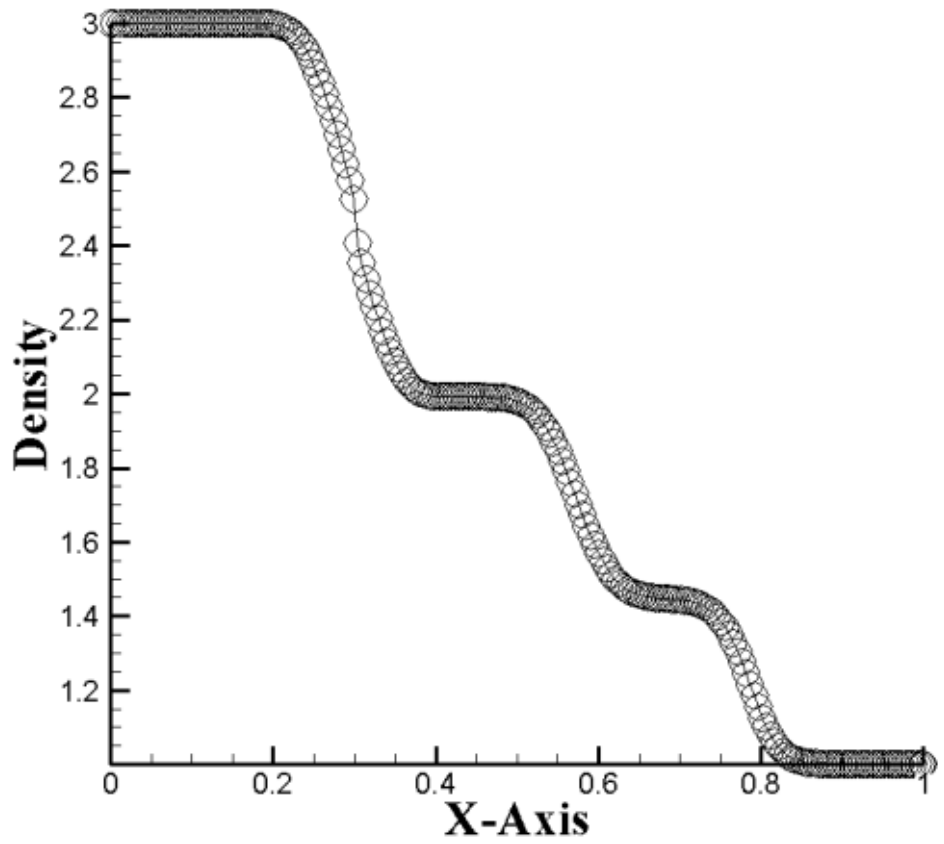

(h) Case 8

Fig. 4 Results of the shock tube test

Fig. 4 shows the effect of changing the terms. The computations of Cases (2), (4),

(5), and (7) diverge, and the results before divergence are provided in Fig. 4. The results reveal that the expansion shock can become considerably serious by decreasing the coefficients of $\Delta U$ in $\delta p_u$ and $\Delta p$ in $\delta U_p$ while increasing those in $\delta p_p$ and $\delta U_u$. Otherwise, the expansion shock is suppressed to increase the coefficients in $\delta p_u$ and $\delta U_p$ and decrease the coefficients in $\delta p_p$ and $\delta U_u$.

According to the preceding discussion, the mechanism of the traditional curing method for Eqs. (22)–(23) and the improvement of Eq. (18) are substantially understood. Eqs. (22)–(23) increase the coefficients in $\delta p_u$ and $\delta U_p$ (see Eqs. (40) and (43)) and cure the expansion shock. Eq. (18) decreases $\delta U_p$ to zero for high Mach number flow, particularly for multi-dimensional calculations when $M \to 1$ but $U \to 0$. Thereafter, the problem of expansion shock deteriorates (see Fig. 3(b)).

## 4. Simultaneous Improvement of Curing Expansion Shock and Shock Instability

Although Eq. (18) worsens the expansion shock, this condition is reasonable and necessary to suppress shock instability [17]. The traditional curing method of Eqs. (22)–(23) only increases the coefficients in $\delta p_u$ and $\delta U_p$ and does not completely utilize the potential to decrease the coefficients in $\delta p_p$ and $\delta U_u$. Therefore, an improved Roe scheme is proposed to simultaneously cure expansion shock and shock instability as follows:

$$\xi = |U|, \tag{57}$$

$$\delta p_p = \text{sign}(U)\min(|U'|, c)\frac{\Delta p}{c}, \tag{58}$$

$$\delta p_u = \max(0, c - |U'|)\rho\Delta U, \tag{59}$$

$$\delta U_p = s_1 s_2 \max\left(0, c - |U'|\right) \frac{\Delta p}{\rho c^2}, \tag{60}$$

$$\delta U_u = \operatorname{sign}(U) \min\left(|U'|, c\right) \frac{\Delta U}{c}, \tag{61}$$

$$|U'| = |U| - \frac{\operatorname{sign}(U+c)\max(0, U_R - U_L) - \operatorname{sign}(U-c)\max(0, U_R - U_L)}{4}. \tag{62}$$

The present scheme is significantly concise and easy to implement; the computational cost only has a negligible increase as well. Compared with the scheme of Eqs. (14)–(16) and (18), only $|U|$ is redefined as $|U'|$ by Eq. (62), which can also be expressed as follows:

$$|U'| = \begin{cases} \min\left(|U_L|, |U_R|\right) & |U| < c \text{ and } U_R > U_L \\ 0 & \text{otherwise} \end{cases}. \tag{63}$$

Thus, the value of $|U|$ is decreased for subsonic expansion flows but still within a reasonable range as given in Eq. (44). Therefore, $\delta p_p$ and $\delta U_u$ decrease and $\delta p_u$ and $\delta U_p$ increase synchronously, which provide sufficient power to cure the expansion shock even $\delta U_p$ is decreased by the functions $s_1$ and $s_2$ in Eq. (19).

Figs. 5 and 6 show the numerical results of the present scheme. Higher-order reconstruction methods [23][24][25] are generally adopted for practical problems; thus, MUSCL reconstruction is also adopted to test the higher-order performance of the improved Eq. (62) or (63). The computational processes are robust and all results are satisfactory, particularly for the supersonic corner test, where the expansion shock and the shock instability are simultaneously cured. No adverse side effects were reported for the improvement.

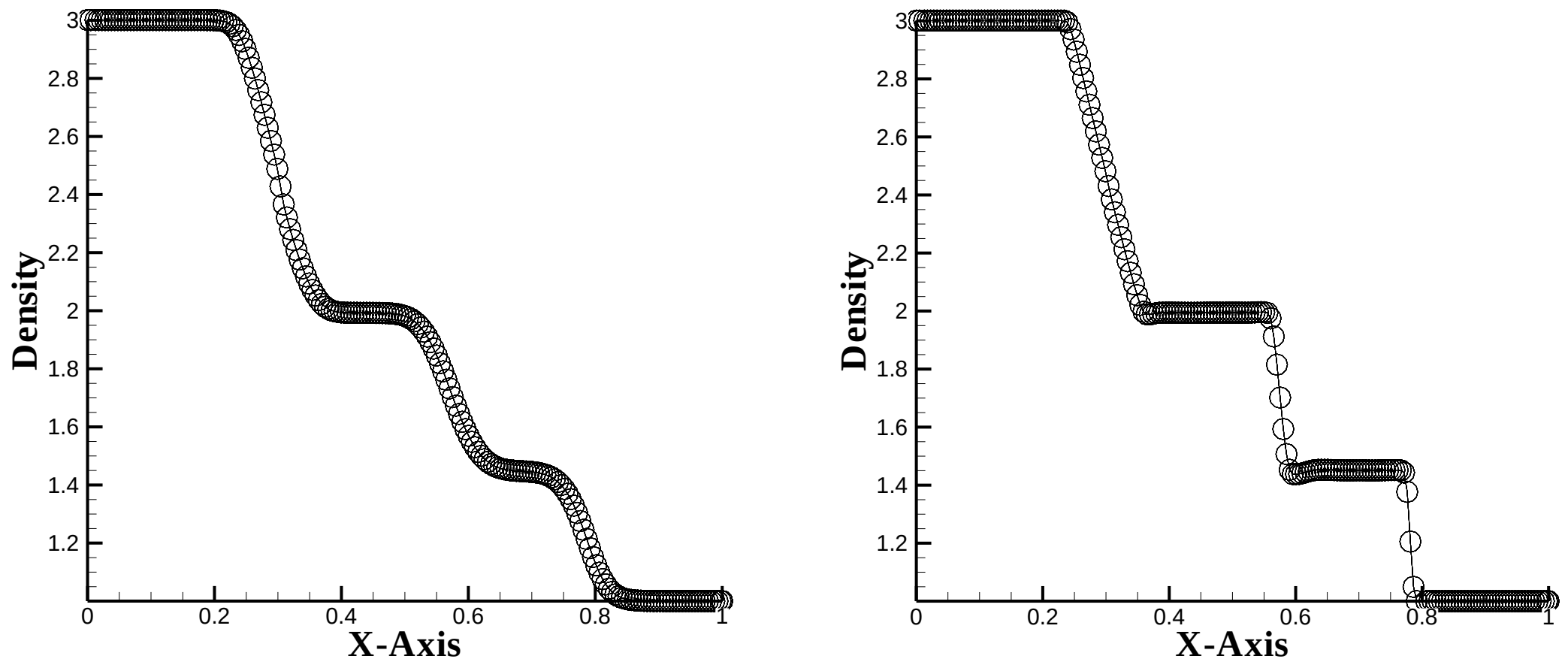

(a) First-order accuracy (b) MUSCL reconstruction

Fig. 5 Results of the shock tube test with the present scheme

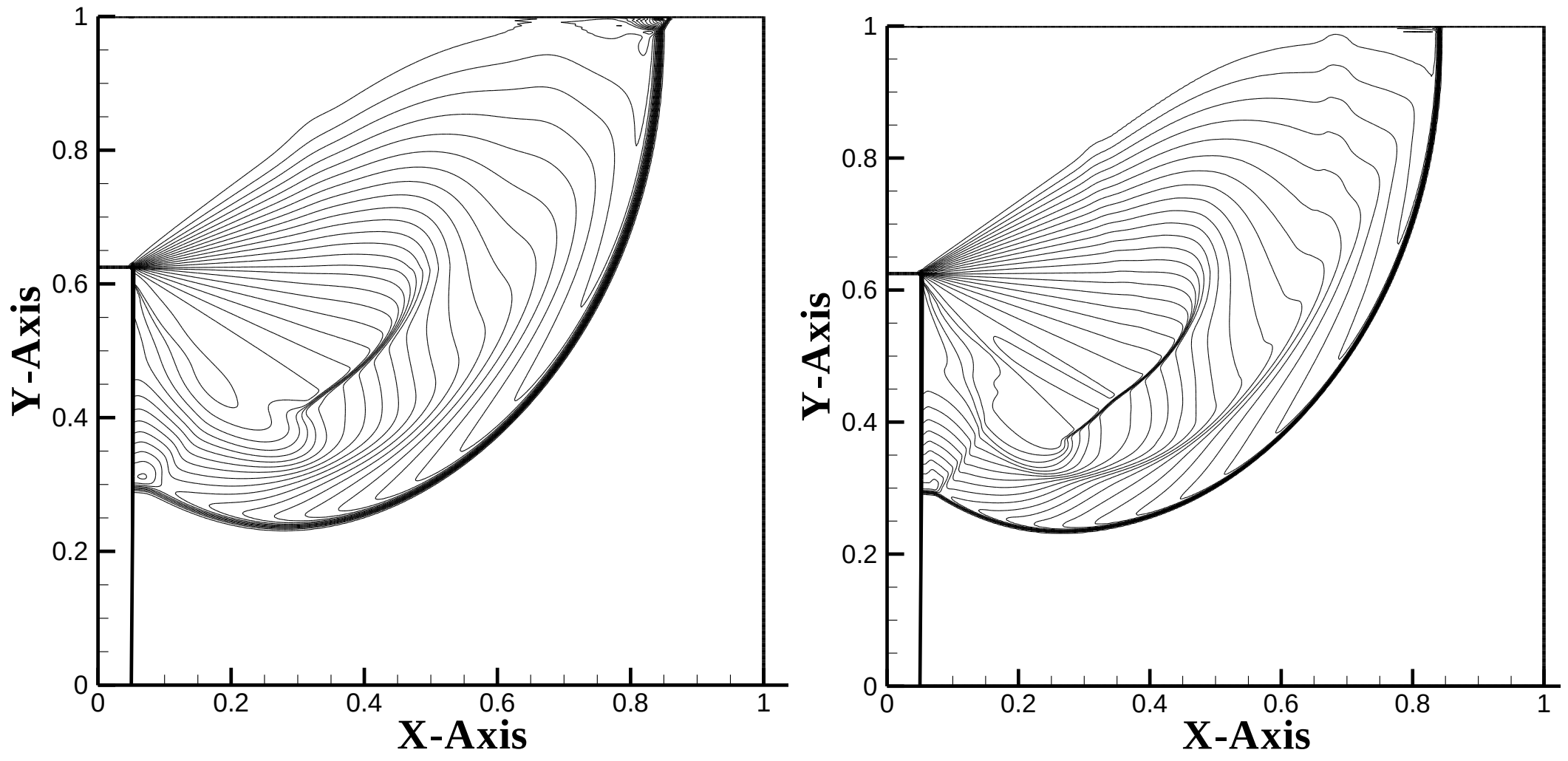

(a) First-order accuracy (b) MUSCL reconstruction

Fig. 6 Results of the supersonic corner test with the present scheme

## 5. Conclusions

The performance of several Roe-type schemes is discussed in terms of expansion shock, and the mechanism of curing expansion shock is analyzed based on the traditional method. Several unfavorable features of the traditional curing method are discovered, and the possible curing mechanism is not completely utilized. Therefore, an improved method is proposed to overcome these problems. The present scheme is substantially concise, easy to implement, and robust with a low computational cost. This scheme is particularly well compatible with the improvement to cure the shock instability. Therefore, the present scheme is simultaneously free from the problems of shock instability and expansion shock without additional expense.

## Acknowledgments

This work is supported by Project 51276092 of the National Natural Science Foundation of China.

## References

[1] P.L. Roe, Approximate Riemann Solvers: Parameter Vectors and Difference Schemes, Journal of Computational Physics 43 (1981) 357-372.

[2] H. Guillard, C. Viozat, On the Behaviour of Upwind Schemes in the Low Mach Number Limit, Computers and Fluids 28 (1999) 63-86.

[3] D.G. Huang, Unified Computation of Flow with Compressible and Incompressible Fluid Based on Roe’s Scheme, Applied Mathematics and Mechanics 27 (2006) 758-763

[4] X.S. Li, C.W. Gu, Mechanism of Roe-type Schemes for All-Speed Flows and Its Application, Computers and Fluids 86 (2013) 56–70.

[5] K. Kitamura, E. Shima, K. Fujimoto, Z.J. Wang. Performance of Low-Dissipation Euler Fluxes and Preconditioned LU-SGS at Low Speeds, Communications in Computational Physics 10 (2011) 90-119.

[6] E. Garnier, M. Mossi, P. Sagaut, P. Comte, and M. Deville. On the Use of Shock-Capturing

Schemes for Large-Eddy Simulation, Journal of Computational Physics 153 (1999) 273-311.

[7] X.S. Li, X.L. Li, All-speed Roe Scheme for the Large Eddy Simulation of Homogeneous Decaying Turbulence, International Journal of Computational Fluid Dynamics, 30 (2016): 69-78.

[8] D.G. Huang, Preconditioned Dual-Time Procedures and its Application to Simulating the Flow with Cavitations, Journal of Computational Physics 223 (2007) 685–689.

[9] J.J. Quirk, A Contribution to the Great Riemann Solver Debate, International Journal for Numerical Methods in Fluids 18 (1994) 555-574.

[10] M.J. Kermani, E.G. Plett, Modified Entropy Correction Formula for the Roe Scheme, AIAA Paper 2001-0083 (2001).

[11] F. Qu, C. Yan, D. Sun, Z. Jiang, A New Roe-type Scheme for All Speeds, Computers & Fluids 121 (2015) 11–25.

[12] S. Kim, C. Kim, O.H. Rho, S.K. Hong, Cures for the Shock Instability: Development of A Shock-Stable Roe Scheme, Journal of Computational Physics 185(2003) 342–374.

[13] Y.X. Ren, A Robust Shock-Capturing Scheme Based on Rotated Riemann Solvers, Computers & Fluids 32 (2003) 1379–1403.

[14] H. Nishikawa, K. Kitamura, Very Simple, Carbuncle-Free, Boundary-Layer-Resolving, Rotated-Hybrid Riemann Solvers, Journal of Computational Physics 227 (2008) 2560–2581.

[15] B. Einfeldt, C. D. Munz, P. L. Roe, and B. Sjögreen, On Godunov–Type Methods near Low Densities, Journal of Computational Physics 92 (1991) 273–295.

[16] M.S. Liou, A Sequel to AUSM, Part II: AUSM+-up for All Speeds, Journal Computational Physics 214 (2006) 137–170.

[17]X.D. Ren, C.W. Gu, and X.S. Li. Role of Momentum Interpolation Mechanism of the Roe Scheme in Shock Instability. arXiv:1509.02776v2 (2015).

[18]X.S. Li, J,Z. Xu and C.W. Gu. Preconditioning Method and Engineering Application of Large Eddy Simulation, Science in China Series G: Physics, Mechanics & Astronomy, 51 (2008) 667-677.

[19]X.S. Li, C.W. Gu, The Momentum Interpolation Method Based on the Time-Marching Algorithm for All-Speed Flows, Journal of Computational Physics 229 (2010) 7806-7818.

[20]A. Pascau, Cell Face Velocity Alternatives in A Structured Colocated Grid for the Unsteady Navier–Stokes Equations, International Journal for Numerical Methods in Fluids 65 (2011) 812–833.

[21]J.M. Weiss, W.A. Smith, Preconditioning Applied to Variable and Const Density Flows, AIAA Journal 33 (1995) 2050-2057.

[22]X.S. Li, Uniform Algorithm for All-Speed Shock-Capturing Schemes, International Journal of Computational Fluid Dynamics 28 (2014) 329–338.

[23]B. Van Leer, Towards the Ultimate Conservative Difference Scheme. V. A Second-Order Sequel to Godunov's Method, Journal of Computational Physics 32 (1979) 101-136.

[24]X. Ren, K. Xu, W. Shyy, and C. Gu, A Multi-Dimensional High-Order Discontinuous Galerkin Method Based on Gas Kinetic Theory for Viscous Flow Computations. Journal of Computational Physics 292 (2015) 176-193.

[25]X. Ren, K. Xu, W. Shyy. A Multi-Dimensional High-Order DG-ALE Method Based on Gas-Kinetic Theory with Application to Oscillating Bodies. Journal of Computational Physics 316 (2016) 700-720.